\documentclass[times,twocolumn,final]{elsarticle}

\usepackage{framed,multirow}

\usepackage{amssymb}
\usepackage{latexsym}

\usepackage{times}
\usepackage{epsfig}
\usepackage{graphicx}
\usepackage{amsmath}
\usepackage{amssymb}

\usepackage{mathrsfs}
\usepackage{multirow}
\usepackage{booktabs}
\usepackage{diagbox}
\usepackage[table,xcdraw]{xcolor}
\usepackage{caption}
\usepackage[TABBOTCAP]{subfigure}
\usepackage{graphicx}
\usepackage{float}
\usepackage{algorithm}
\usepackage{algorithmicx}
\usepackage{algpseudocode}

\usepackage{url}
\usepackage{xcolor}
\usepackage{hyperref}

\definecolor{newcolor}{rgb}{.8,.349,.1}

\begin{document}

% \verso{Donglin Di \textit{et~al.}}

\begin{frontmatter}

\title{Hypergraph Learning for Identification of COVID-19 with CT Imaging}

\author[1]{Donglin Di \fnref{DDL}}
\fntext[DDL]{D. Di, F. Shi, F. Yan, L. Xia, Z. Mo, Z. Ding, F. Shan contributed equally to this work.}
% \ead{donglin.ddl@gmail.com}

\author[2]{Feng Shi \fnref{DDL}} % \fnref{FShi}}
% \fntext[FShi]{F. Shi, Y. Wei, Y. Shao, M. Han, Yaozong Gao and D. Shen are with the Department of Research and Development, Shanghai United Imaging Intelligence Co., Ltd., Shanghai 200232, China. \textit{(email: {feng.shi, ying.wei, ying.shao, miaofi.han, yaozong.gao}@united-imaging.com)}}
% % \ead{feng.shi@united-imaging.com}

\author[3]{Fuhua Yan \fnref{DDL}} % \fnref{FYan}}
% \fntext[FYan]{F. Yan is with the Department of Radiology, Ruijin Hospital, Shanghai Jiao Tong University School of Medicine, Shanghai, China. \textit{(email: yfh11655@rjh.com.cn)}}
% % \ead{yfh11655@rjh.com.cn}

\author[4]{Liming Xia \fnref{DDL}} % \fnref{LXia}}
% \fntext[LXia]{L. Xia is with the Department of Radiology, Tongji Hospital, Tongji Medical College, Huazhong University of Science and Technology, Wuhan, Hubei, China. \textit{(email: xialiming2017@outlook.com)}}
% % \ead{xialiming2017@outlook.com}

\author[5]{Zhanhao Mo \fnref{DDL}} % \fnref{ZMo}}
% \fntext[ZMo]{Z. Mo and H. Sui are with the Department of Radiology, China-Japan Union Hospital of Jilin University, Changchun, China. \textit{(email: mozhanhao@jlu.edu.cn, suihe910402@126.com)}}
% % \ead{mozhanhao@jlu.edu.cn}

\author[6]{Zhongxiang Ding \fnref{DDL}} % \fnref{ZDing}}
% \fntext[ZDing]{Z. Ding is with the Department of Radiology, Hangzhou First People’s Hospital, Zhejiang University School of Medicine, Zhejiang, China. \textit{(hangzhoudzx73@126.com)}}
% % \ead{hangzhoudzx73@126.com}

\author[7]{Fei Shan \fnref{DDL}} % \fnref{FShan}}
% \fntext[FShan]{F. Shan is with the Department of Radiology, Shanghai Public Health Clinical Center, Fudan University, Shanghai, China. \textit{(shanfei\_2901@163.com)}}
% % \ead{shanfei_2901@163.com}

\author[1]{Shengrui Li}
% \ead{lisr17@mails.tsinghua.edu.cn}

\author[2]{Ying Wei}
% \ead{ying.wei@united-imaging.com}

\author[2]{Ying Shao}
% \ead{ying.shao@united-imaging.com}

\author[2]{Miaofei Han}
% \ead{miaofei.han@united-imaging.com}

\author[2]{Yaozong Gao}
% \ead{yaozong.gao@united-imaging.com}

\author[5]{He Sui}
% \ead{suihe910402@126.com}

\author[1]{Yue Gao \corref{cor1}}
% \ead{gaoyue@tsinghua.edu.cn}

\author[2]{Dinggang Shen \corref{cor1}}

\cortext[cor1]{Correspondence authors: Yue Gao (gaoyue@tsinghua.edu.cn), and Dinggang Shen (dinggang.shen@gmail.com)}
% \ead{dinggang.shen@gmail.com}

\address[1]{BNRist, THUIBCS, KLISS, School of Software, Tsinghua University, Beijing, China.}
\address[2]{Department of Research and Development, Shanghai United Imaging Intelligence Co., Ltd., Shanghai, China.}
\address[3]{Department of Radiology, Ruijin Hospital, Shanghai Jiao Tong University School of Medicine, Shanghai, China.}
\address[4]{Department of Radiology, Tongji Hospital, Tongji Medical College, Huazhong University of Science and Technology, Wuhan, Hubei, China.}
\address[5]{Department of Radiology, China-Japan Union Hospital of Jilin University, Changchun, China.}
\address[6]{Department of Radiology, Hangzhou First People’s Hospital, Zhejiang University School of Medicine, Zhejiang, China.}
\address[7]{Department of Radiology, Shanghai Public Health Clinical Center, Fudan University, Shanghai, China.}

% \received{27 April 2020}
% \finalform{27 April 2020}
% \accepted{27 April 2020}
% \availableonline{27 April 2020}
% \communicated{A. Editor}

\begin{abstract}
The coronavirus disease, named COVID-19, has become the largest global public health crisis since it started in early 2020.
CT imaging has been used as a complementary tool to assist early screening, especially for the rapid identification of COVID-19 cases from community acquired pneumonia (CAP) cases.
The main challenge in early screening is how to model the confusing cases in the COVID-19 and CAP groups, with very similar clinical manifestations and imaging features.
To tackle this challenge, we propose an Uncertainty Vertex-weighted Hypergraph Learning (UVHL) method to identify COVID-19 from CAP using CT images.
In particular, multiple types of features (including regional features and radiomics features) are first extracted from CT image for each case.
Then, the relationship among different cases is formulated by a hypergraph structure, with each case represented as a vertex in the hypergraph.
The uncertainty of each vertex is further computed with an uncertainty score measurement and used as a weight in the hypergraph.
Finally, a learning process of the vertex-weighted hypergraph is used to predict whether a new testing case belongs to COVID-19 or not.
Experiments on a large multi-center pneumonia dataset, consisting of 2,148 COVID-19 cases and 1,182 CAP cases from five hospitals, are conducted to evaluate the performance of the proposed method.
Results demonstrate the effectiveness and robustness of our proposed method on the identification of COVID-19 in comparison to state-of-the-art methods.
\end{abstract}


\end{frontmatter}

% \makeatletter
\def\eg{\emph{e.g.}} \def\Eg{\emph{E.g}}
\def\ie{\emph{i.e.}} \def\Ie{\emph{I.e.}}
\def\cf{\emph{c.f.}} \def\Cf{\emph{C.f.}}
\def\etc{\emph{etc. }} \def\vs{\emph{vs.}}
\def\aka{\emph{a.k.a.}}
\def\wrt{w.r.t.} \def\dof{d.o.f.}
\def\etal{\emph{et al.}}
% \makeatother

\section{Introduction}

The coronavirus disease pandemic, named COVID-19, has become the largest global public health crisis since in started it early of 2020.
COVID-19 was caused by a kind of savagely contagious virus, and could lead to acute respiratory distress and multiple organ failure \citep{li2020artificial, chen2020epidemiological, li2020early, wang2020clinical, holshue2020first}. 

The latest guideline, published by the Chinese government (the trial sixth version) \citep{generaloffice}, declares that the diagnosis of COVID-19 must be confirmed by the reverse transcription polymerase chain reaction (RT-PCR) or gene sequencing for respiratory or blood specimens.
Recent studies \citep{fang2020sensitivity, gozes2020rapid, xie2020chest} have investigated the sensitivity of non-contrast chest CT, and demonstrated that, recognizing either diffusion or focal ground-glass opacities as the disease characteristics in CT is a reliable and efficient approach.
More specifically, the bilateral and peripheral ground-class and consolidative pulmonary opacities in CT are the typical features of COVID-19 symptoms, and the greater severity of the disease with increasing time from onset symptoms shows larger lung involvement and more linear opacities, \textit{a.k.a.} the ``crazy-paving'' pattern and the ``reverse halo'' sign \citep{xie2020chest, bernheim2020chest}.

To reduce the workload in diagnosing COVID-19, plenty of machine learning-based studies have been conducted \citep{gozes2020rapid, li2020artificial, narin2020automatic, zhang2020covid, shan+2020lung}.
However, there are still two major challenges:
\textbf{1)} Noisy data, due to the large variations of data collected in an emergent situation, such as using different reconstruction kernels and CT manufactures, along with possible patient movements;
\textbf{2)} Confusing cases, due to similar radiological appearance of COVID-19 and other pneumonia, especially in the early stage.
Therefore, how to handle these challenges is the key for successful identification of COVID-19 from CAP.

Accordingly, in this work, we propose an uncertainty based learning framework, called Uncertainty Vertex-weighted Hypergraph Learning (UVHL), to identify COVID-19 from CAP with CT images.
The most essential task is to exploit the latent relationship among various COVID-19 cases and CAP cases, and then make a prediction for a new testing case, \ie, whether belonging to COVID-19 or not.
The proposed framework employs a vertex-weighted hypergraph structure to formulate data correlation among different cases.
The module of ``uncertainty score measurement'' is used to generate two metrics, \ie, \textbf{1)} noisy data aleatoric uncertainty and \textbf{2)} the model’s inability epistemic uncertainty.
Then, the proposed UVHL conducts learning on the hypergraph structure to make a prediction for the new testing case, by simultaneously \textbf{a)} incorporating the uncertainty values of measured data to relieve the misleading patterns from noisy low-quality data and \textbf{b)} allocating more attention to the nodes distributing around the classifying interface in the latent representation space.
Another advantage of the proposed framework is its flexibility in utilizing multi-modal data/features when available.
We apply our proposed method to a large dataset, with 2,148 COVID-19 cases and 1,182 CAP cases.
The experimental results show that our proposed method can achieve a satisfactory accuracy of 90\% for identification of COVID-19 from CAP.

The main contributions of this paper are summarized as follows:
\begin{itemize}
    \item We propose to formulate data correlation among all COVID-19 and CAP cases using hypergraph, for exploring high-order relationship using multi-type CT features (such as regional features and radiomics features).
    \item We propose an uncertainty vertex-weighted strategy to relieve the influence of noisy (CT) data collected from suspected COVID-19 patients in emergent situation.
    \item We have demonstrated better performance in the task of identifying COVID-19 from CAP, and have also shown how different types of CT features perform in this task.
\end{itemize}

% ---------------------- Figure -----------------------
\begin{figure}
    \centering
    \includegraphics[width=\linewidth]{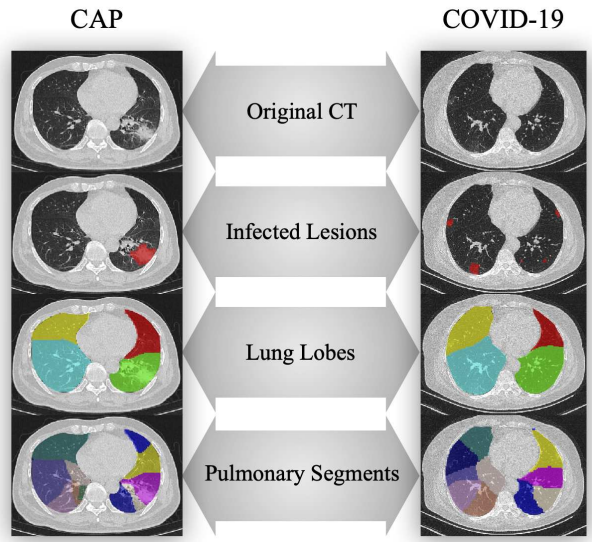}
    \caption{\label{fig:covid_compare} Illustration of lung CT image, infection, lung lobes, and pulmonary segments on a CAP case (left) and a COVID-19 case (right).}
    \vspace{-0.4cm}
\end{figure}
% ---------------------- Figure -----------------------

\section{Related Work}
In this section, we briefly review recent works on diagnosing COVID-19 and introduce current studies on hypergraph learning.

\subsection{AI-based COVID-19 Diagnosis}
As introduced in \citep{zu2020coronavirus}, COVID-19 patients could be divided into mild, moderate, severe and critically ill stages, according to the severity of disease development.
In the mild stage, the pneumonia symptom is difficult to be observed from CT images for a suspected patient.
With the development of the disease, ground-glass opacity (GGO), increased crazy-paving pattern, and consolidation can be observed \citep{li2020coronavirus}.
When it becomes a serious situation, the symptom will deteriorate and also the gradual resolution of consolidation could be observed in CT images.

In the very early studies, several statistics-based methods \citep{chen2020epidemiological, li2020early, wang2020clinical} are proposed to develop automatic detection and patient monitoring methods for diagnosis of COVID-19.
However, only simple data statistics is employed in these methods, which limits the capability of diagnosing suspected patients when facing the challenge of noisy data and confusing cases.

To further improve the performance, a group of AI-based methods \citep{narin2020automatic, shan+2020lung, gozes2020rapid} are proposed in the following.
In \cite{bernheim2020chest, shan+2020lung, tang2020severity}, reliable representations from CT are learned to represent the symptom of COVID-19.
The co-relationship between chest CT and RT-PCR testing has also been investigated in COVID-19 \citep{ai2020correlation, fang2020sensitivity, xie2020chest}.
\cite{gozes2020rapid} introduce an AI-based automatic CT image analysis tool for detection, quantification, and tracking of coronavirus.

% Furthermore, to detect and monitor COVID-19 based on CT images, several methods have been introduced in \cite{narin2020automatic, zhang2020covid}.
% According to recent works \cite{li2020coronavirus, narin2020automatic}, the chest CT is a routine diagnostic tool for pneumonia.
% Especially with thin slices \cite{wong2003thin}, they are very useful in detecting typical radiographic features of COVID-19.

Although there have been plenty of works on AI-assisted COVID-19 diagnosis tools, the identification of COVID-19 from CAP has not fully investigated, which has become an important issue recently.
In this task, \cite{wang2020deep} propose to classify the patches of infected lesions into COVID-19 or typical viral pneumonia using the modified and fine-tuned Inception migration-learning model with the pre-trained weights, in which the infection patches need to be manually labeled.
Another issue is the correlation among the COVID-19 cases and the CAP cases, which is important to identify the category of a new testing case.

\subsection{Preliminary on Hypergraph Learning}
Hypergraph learning has been widely applied in many tasks, such as identifying non-random structure in structural connectivity of the cortical microcircuits \citep{dotko2016topological}, identifying high-order brain connectome biomarkers for disease diagnosis \citep{zu2016identifying}, and studying the co-relationships between functional and structural connectome data \citep{munsell2016identifying}.
Hypergraph learning was first introduced in \citep{zhou2007learning}, in which each node represents one case, each hyperedge captures the correlation between each pair of nodes, and the learning process is conducted on a hypergraph as a propagation process.
By this method, the transductive inference on hypergraph aims to minimize the label differences between vertices that are connected by more and stronger hyperedges.
Then, the hypergraph learning is conducted as a label propagation process on the hypergraph to obtain the label projection matrix \citep{liu2017view}, or as a spectral clustering \citep{li2017inhomogeneous}.

Other applications of hypergraph learning include video object segmentation \citep{huang2009video}, images ranking \citep{huang2010image}, and landmark retrieval \citep{zhu2015content}.
Hypergraph learning has the advantage of modeling high-order correlation modeling, but the reliability of different vertices on the hypergraph, also important to conduct accurate learning, has not been well investigated.

% In our work, we make use of the hypergraph structure to modeling the high-order representations of COVID-19 cases based on the patients' image features from CT as well as other multi-modal fruitful informative data, and also propose to measure the uncertainty further to distinguish the pneumonia patients.

\section{Materials and Preprocessing}
In this section, we first introduce materials used in this work and image preprocessing steps.
Then, multi-type features, including regional features and radiomics features from CT images are extracted.

% ---------------------- Figure -----------------------
\begin{figure*}
    \begin{center}
    \includegraphics[width=\linewidth]{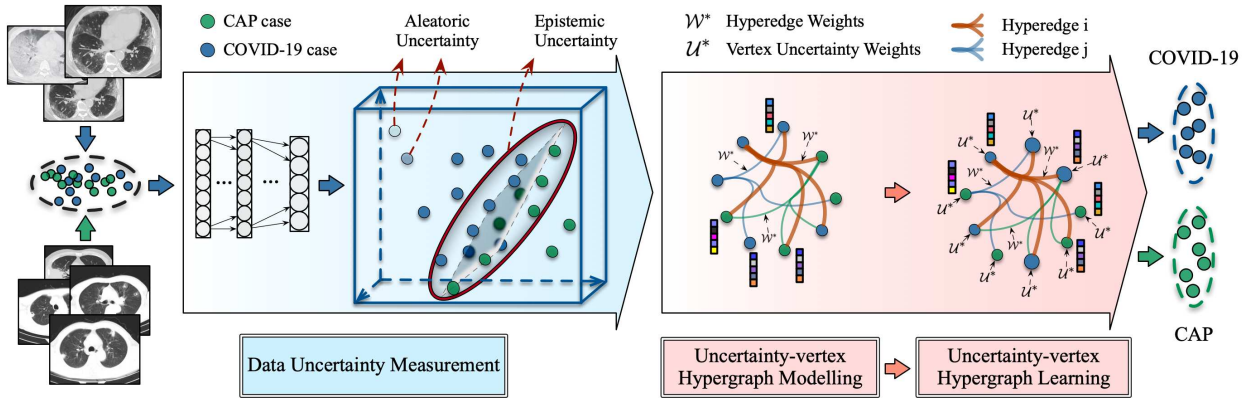}
    \end{center}
    \caption{\label{fig:framework} Illustration of our proposed Uncertainty Vertex-weighted Hypergraph Learning (UVHL) method for COVID-19 identification. %We generate uncertainty scores and coarse classification in the first stage. Then, the vertex-weighted hypergraph network can focus on learning the high-order representations for the more weighted vertices without being influenced by abnormal noisy data. Finally, our framework classifies could provide more precise classification.
    }
\end{figure*}
% ---------------------- Figure -----------------------

\subsection{Dataset}
In this study, a total of 3,330 CT images were collected, including 2,148 from COVID-19 patients and the rest 1,182 from CAP patients.
These images were provided by the Ruijin Hospital of Shanghai Jiao Tong University, Tongji Hospital of Huazhong University of Science and Technology, China-Japan Union Hospital of Jilin University, Hangzhou First People’s Hospital of Zhejiang University, and Shanghai Public Health Clinical Center of Fudan University. 
All the COVID-19 cases were confirmed as positive by RT-PCR and acquired from Jan. 9, 2020 to Feb. 14, 2020.
CAP images were obtained from Jul. 30, 2018 to Feb. 22, 2020.
The CT scanners used in this study include uCT 780 from UIH, Optima CT520, Discovery CT750, LightSpeed 16 from GE, Aquilion ONE from Toshiba, SOMATOMForce from Siemens, and SCENARIA from Hitachi.
The CT protocol here includes: 120 kV, reconstructed CT thickness ranging from 0.625 to 2mm, and breath-hold at full inspiration.
All images were de-identified before sending for analysis.
This study was approved by the Institutional Review Board of participating institutes.
Written informed consent was waived due to retrospective nature of the study.

\subsection{Preprocessing}
\label{sec:preprocess}
In this study, both regional and radiomics features are extracted from CT image for each patient.
More specifically, we first perform segmentation of left / right lung, 5 lung lobes, and 18 pulmonary segments, as well as infected lesions by deep learning based network, \ie, VB-Net, in a portal software \citep{shan+2020lung}, for each CT image.

To generate regional features, we calculate a dimension of $\mathbb{R}^{96}$ features for each patient, including histogram distribution, infected lesion counting numbers, the mean and variance grey values of lesion area, lesion surface area, and additional density and mass features, \etc
To generate radiomics features, radiomics computation is performed on the infected lesions and a dimension of $\mathbb{R}^{93}$ for each patient is extracted, including the first-order intensity statistics and texture features such as gray level co-occurrence matrix \citep{shi2020large}.
With the information on age and sex also included, the representations for each patient can be concatenated as $\mathbf{x}\in \mathbb{R}^{191}$ overall.

\section{The Method}
In this section, we introduce our proposed Uncertainty Vertex-weighted Hypergraph Learning (UVHL) method for COVID-19 identification.
Figure~\ref{fig:framework} shows in the framework of our proposed method, which is composed of three steps, \ie, \textbf{1)} ``Data Uncertainty Measurement'', \textbf{2)} ``Uncertainty-vertex Hypergraph modeling'' and \textbf{3)} ``Uncertainty-vertex Hypergraph Learning'', respectively.

\subsection{Data Uncertainty Measurement}
\label{subsec: uncertainty_measure}
As introduced before, the data quality may suffer from the unstable, noisy nature caused in the emergent situation.
To overcall this limitation, it is important to identify the reliability of different cases during the learning processing.
In this step, a data uncertainty measurement process is conducted to generate uncertainty scores for all cases used in the learning processing.
Here, two types of uncertainty factors are calculated in our method.

\begin{enumerate}
    \item[\textit{a.}] \textit{Aleatoric Uncertainty.} The data is abnormal, noisy or collected by mistake with low quality.
    \item[\textit{b.}] \textit{Epistemic Uncertainty.} The features of these cases lie around the decision boundary that makes the distinguishing model under a serious challenge.
\end{enumerate}
We will introduce how to calculate these uncertainty scores in details as below.

\subsubsection{Aleatoric Uncertainty}

The aleatoric uncertainty represents the measure of the quality measure of noisy data, and is based on the comparison of data distributions.
The objective is to estimate $\Theta$ that minimizes the Kullback-Leibler (KL) divergence between true distribution $P_{D}(x)$ and predicted distribution $P_{\Theta}(x)$ over $N$ training samples:
\begin{equation}
    \hat{\Theta}=\underset{\Theta}{\arg \min } \frac{1}{N} \sum_{i=1}^N D_{K L}\left(P_{D}(x) \| P_{\Theta}(x)\right)
\end{equation}

Hence, the loss function can be defined as KL-Divergence: $L(\Theta)=L_{KL}(\Theta)$, which is minimized during the training process.
In detail, the loss for a single case can be calculated as Eq.~\ref{eq:kl_loss}:
\begin{equation}
\begin{aligned}
L(\Theta) &=
D_{K L}\left(P_{D}(\mathbf{x}) \| P_{\Theta}(\mathbf{x})\right) \\
&=\int P_{D}(\mathbf{x}) \log P_{D}(\mathbf{x}) \mathrm{d} \mathbf{x}-\int P_{D}(\mathbf{x}) \log P_{\Theta}(\mathbf{x}) \mathrm{d} \mathbf{x} \\
&=\!\frac{\mathbb{CE}(\mathbf{y},f_{\Theta}(\mathbf{x}))}{2 {\sigma^{2}_{\Theta}(\mathbf{x})}}\!+\!\frac{\log \left({\sigma^{2}_{\Theta}(\mathbf{x})}\right)}{2}\!+\!\frac{\log (2 \pi)}{2}\!-\!H(P_{D}(\mathbf{x}))
\label{eq:kl_loss}
\end{aligned}
\end{equation}
where $\mathbb{CE}$ denotes the \textit{Cross-Entropy} function, $\mathbf{x} \in \mathbb{R}^{191}$ denotes the feature vector of each patient, $\mathbf{y} \in \mathbb{R}^{2}$ is the label, and $f_{\Theta}: \mathbb{R}^{191} \mapsto \mathbb{R}^{2}$ represents the network with \textit{softmax} function as the last layer that maps features to the corresponding binary prediction.
$H({P_{D}(\mathbf{x})})$ stands for the entropy of ${P_{D}(\mathbf{x})}$.
$\sigma^{2}_{\Theta}$ denotes the predicted variance.
To avoid the potential division by zero, we replace $\log {\sigma}^2_{\Theta}(\mathbf{x})$ by $\alpha_{\Theta}(\mathbf{x})$.
Therefore, $\alpha_{\Theta}: \mathbb{R}^{191} \mapsto \mathbb{R}^{1}$ can be used to predict the uncertainty score for each case.

Note that $\log {(2 \pi)}/2$ and $H\left(P_{D}(\mathbf{x})\right)$ are redundant for optimization.
Therefore, for $N$ samples, we can rewrite the loss function as Eq.~\ref{eq:n_sample_func}:
\begin{equation}
L(\Theta)\!=\!\frac{1}{N} \sum_{i}^N \left(\!\frac{1}{2} exp( {-{\mathbf{\alpha}_{\Theta}(\mathbf{x}_i)}})\mathbb{CE}(\mathbf{y}_i,f_{\Theta}(\mathbf{x}_i))\!+\!\frac{1}{2} \mathbf{\alpha}_{\Theta}(\mathbf{x}_i)\!\right)
\label{eq:n_sample_func}
\end{equation}

If the \textit{Cross-Entropy} between the predicted ${y}_{\Theta}(\mathbf{x}_i)$ and true label $\mathbf{y}_i$ is quite large, the model tends to predict a higher $\mathbf{\alpha}_{\Theta}(\mathbf{x_i})$ to make inputs with high uncertainty having a smaller effect on the loss.
This allows the network to learn to attenuate the effect from erroneous labels, thus becoming more robust to noisy data.
In our task, we denote $\mathcal{A}_\Theta(\mathbf{x}_i)$ as aleatoric uncertainty to identify low quality data, as defined in Eq.~\ref{aleatoric_uncertainty}:
\begin{equation}
    \label{aleatoric_uncertainty}
    \mathcal{A}_\Theta(\mathbf{x}_i)=\sigma^2_\Theta(\mathbf{x}_i)=exp({\mathbf{\alpha}_{\Theta}(\mathbf{x_i})})
\end{equation}

\subsubsection{Epistemic Uncertainty}
Epistemic uncertainty refers to the model's inability for accurate and precise prediction.
To compute this measurement, we use the dropout variation inference, which is a widely adopted practical approach for approximate inference \citep{gal2016dropout}.
The Monte Carlo estimation method is referred as MC dropout.
Our approximate predictive distribution is given by Eq.~\ref{eq:mc_dropout}:
\begin{equation}
q\left(\mathbf{y}^{*} | \mathbf{x}^{*}\right)=\int p\left(\mathbf{y}^{*} | \mathbf{x}^{*}, \boldsymbol{\omega}\right) q(\boldsymbol{\omega}) \mathrm{d} \boldsymbol{\omega}
\label{eq:mc_dropout}
\end{equation}
where $\boldsymbol{\omega}=\left\{\mathbf{W}_{i}\right\}_{i=1}^{L}$ is a set of random variables for a model with $L$ layers.
$\mathbf{x}^*$ and $\mathbf{y}^*$ denote the input and the corresponding output of any MC dropout model, respectively.
The effect of our MC dropout can be attributed to impose a Gaussian distribution on each layer during the test stage.
In detail, the multi-layer perception neural network (MLP) model can be trained with dropout. But different from the conventional settings, these dropout layers are kept open during the testing stage.
Each case is predicted for $K$ times, and the epistemic uncertainty for this case can be calculated using the variance of these $K$ values.

Therefore, the predicted result for one case can be obtained by Eq.~\ref{eq:pred_soft}:
\begin{equation}
\mathbf{E}_{q\left(\mathbf{y}^{*} | \mathbf{x}^{*}\right)}\left(\mathbf{y}^{*}\right) \approx \frac{1}{K} \sum_{k=1}^{K} \widehat{\mathbf{y}}^{*}\left(\mathbf{x}^{*}, \boldsymbol{\omega}^k \right)
\label{eq:pred_soft}
\end{equation}
or more specifically by Eq.~\ref{eq:pred_soft_rewrite} in our task:
\begin{equation}
\mathbf{E}(f_{\widehat{\Theta}}(\mathbf{x}_i))\approx \frac{1}{K} \sum_{k=1}^{K} f_{\widehat{\Theta}(\boldsymbol{\omega}^k)}(\mathbf{x}_i)
\label{eq:pred_soft_rewrite}
\end{equation}

Combined with aleatoric uncertainty introduced before, the epistemic uncertainty can be approximated as~\citep{kendall2017uncertainties} in Eq.~\ref{equ:epistemic}:

\begin{equation}
\begin{aligned}
{\mathscr{E}}(f_{\widehat{\Theta}}(\mathbf{x}_i)) \approx & {\mathcal{A}_{\widehat{\Theta}}}(\mathbf{x}_i)+\frac{1}{K} \sum_{k=1}^{K} f_{\widehat{\Theta}(\boldsymbol{\omega}^k)}(\mathbf{x}_i)^{T} f_{\widehat{\Theta}(\boldsymbol{\omega}^k)}(\mathbf{x}_i)\\
& -\mathbf{E}(f_{\widehat{\Theta}(\boldsymbol{\omega}^k)}(\mathbf{x}_i))^T\mathbf{E}(f_{\widehat{\Theta}(\boldsymbol{\omega}^k)}(\mathbf{x}_i))
\label{equ:epistemic}
\end{aligned}
\end{equation}
where $i$ denotes the $i_{th}$ sample and $k$ denotes the $k_{th}$ test with dropout. 

Note that ${\mathscr{E}}(f_{\widehat{\Theta}}(\mathbf{x}_i)) \approx {\mathcal{A}_{\widehat{\Theta}}}(\mathbf{x}_i)$ (epistemic uncertainty) is mainly composed of aleatoric uncertainty.
Consequently, when ${\mathscr{E}}(f_{\widehat{\Theta}}(\mathbf{x}_i))$ gets higher, it mainly represents lower data quality instead of the limitation on classification capability.

To normalize the epistemic uncertainty ${\mathscr{E}}(f_{\widehat{\Theta}}(\mathbf{x}_i))$, its mean and standard deviation in the whole dataset can be calculated as $\mathbf{\mu}_{e}, \mathbf{s}_{e}$.
Then, \textit{sigmoid} function $\sigma(\cdot)$ is adopted to ensure the uncertainty score ranging from 0 to 1.
$\lambda$ is an adjustable parameter, to make different uncertainty cases more distinctive.
If the $\lambda$ is set to positive, the cases with the high uncertainty score will be adjusted higher, the cases with the low uncertainty score will be lower, and vice versa.
Weights of all data are shown in Eq.~\ref{eq:weight}:
\begin{equation}
    \mathcal{U}_i = 
    \sigma \left( \lambda \frac{{\mathscr{E}}(f_{\widehat{\Theta}}(\mathbf{x}_i))-\mathbf{\mu}_{e}}{\mathbf{s}_{e}}\right)
\label{eq:weight}
\end{equation}
In the end of this step, by leveraging the uncertainty, the quality of data is measured and also the weighted vertices are generated accordingly. 

\subsection{Uncertainty-vertex Hypergraph Construction}
To identify the COVID-19 cases, it is important to exploit the data correlation.
Here, the hypergraph structure is employed to model the relationship among the known training COVID-19 cases, the known training CAP cases, and the unknown testing cases.  

In the hypergraph $\mathcal{P} = \langle \mathcal{F}, L, \mathcal{U} \rangle$, each vertex denotes one case, and there are totally $n$ vertices according to the number of cases involved.
Given the two types of features, \ie, the regional features and radiomics features, two groups of hyperedges are generated to build the connections among these cases.
For the regional features, each time one vertex (case) is selected as the centroid, and its $k$ nearest neighbors (cases) are selected to be connected by one hyperedge.
This process repeats until all vertices have been selected once.
Then, a group of hyperedges based on the regional feature can be generated.
The same process is performed for the radiomics feature, which generates another group of hyperedges.
These two groups of hyperedges are concatenated to build the final hypergraph.

Different from conventional hypergraph, the uncertainty-vertex hypergraph not only cares about features $\mathcal{F}$ and the label $L$ of each vertex, but also considers the uncertainty $\mathcal{U}$ of each vertex.
In this way, these more reliable vertices could contribute more during the learning process, and vice versa.
Here, $\mathcal{V}$ is the vertex set, $\mathcal{E}$ is the hyperedges set, and $\mathbf{W}$ is the pre-defined matrix of hyperedge weights.
Besides these, $\mathbf{U}$ denotes the uncertainty matrix for all the vertices.
Therefore, our uncertainty-vertex hypergraph can be written as $\mathcal{G} = \langle \mathcal{V}, \mathcal{E}, \mathbf{W}, \mathbf{U} \rangle$.
Leveraging vertex weights $\mathbf{U}$, an incidence matrix $\mathbf{H}$ is then generated to represent the relationship among different vertices.
\begin{equation}
\mathbf{H}(v_i, e_p) = \left\{\begin{matrix}
\mathcal{U}_i, & v_i \in e_p\\ 
0, & v_i \notin e_p
\end{matrix}\right.
\end{equation}

In the end of this stage, the uncertainty vertex-weighted hypergraph is constructed to represent the correlation among all cases.

% ---------------------- Figure -----------------------
\begin{figure}
    \centering
    \includegraphics[width=\linewidth]{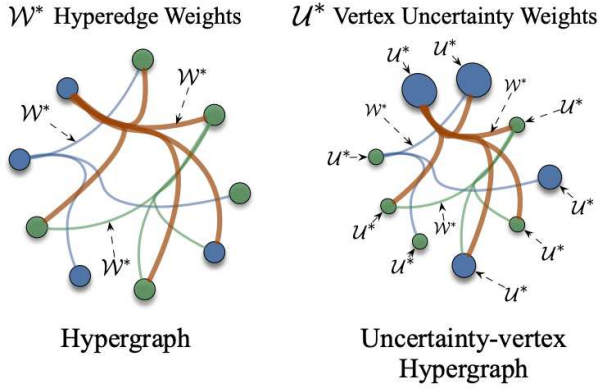}
    \caption{\label{fig:hl_compare} Besides the hyperedge weights, the uncertainty-vertex hypergraph contains the uncertainty score of each vertex.}
    \vspace{-0.4cm}
\end{figure}
% ---------------------- Figure -----------------------

\subsection{Uncertainty-vertex Hypergraph Learning}
As shown in Fig.~\ref{fig:hl_compare}, compared with the conventional hypergraph learning method, the proposed UVHL structure considers the uncertainty of each vertex individually and the learning process is conducted on an unequal space.
The learning task on the uncertainty-vertex hypergraph can be formulated as:
\begin{equation}
\mathcal{Q}_{\mathbf{U}}(\mathbf{F}) = \arg \min _{\mathbf{F}}\left\{\Omega(\mathbf{F})+\lambda \mathcal{R}_{emp}(\mathbf{F})\right\}
\end{equation}

More specifically, the smoothness regularizer function $\Omega(\cdot)$ and the empirical loss term $\mathcal{R}_{e m p}(\cdot)$ can be, respectively, rewritten as follows:
\begin{equation}
\begin{aligned}
\Omega (\mathbf{F}, \mathcal{V}, \mathbf{U}, \mathcal{E}, \mathbf{W}) &= tr(\mathbf{F}^\top (\mathbf{U}^\top - \mathbf{U}^\top \mathbf{\Theta}_{\mathbf{U}} \mathbf{U}) \mathbf{F}) \\
\mathcal{R}_{emp} (\mathbf{F}, \mathbf{U}) &= \sum_{k = 1}^K \begin{Vmatrix} \mathbf{F}(:, k) - \mathbf{Y}(:, k) \end{Vmatrix}^2
\end{aligned}
\end{equation}
where $\mathbf{F}(:, k)$ is the $k_{th}$ column of $\mathbf{F}$ and $\Theta_{\mathbf{U}}=\mathbf{D}_{v}^{-\frac{1}{2}} \mathbf{H} \mathbf{W} \mathbf{D}_{e}^{-1} \mathbf{H}^{\mathrm{T}} \mathbf{D}_{v}^{-\frac{1}{2}}$.
The uncertainty vertex-weighted hypergraph loss function $\mathcal{R}_{e m p}(\cdot)$ can be further rewritten as:
\begin{equation}
\begin{aligned}
\mathcal{R}_{emp} (\mathbf{F}, \mathbf{U}) = &tr(\mathbf{F}^\top \mathbf{U}^\top \mathbf{U} \mathbf{F} + \mathbf{Y}^\top \mathbf{U}^\top \mathbf{U} \mathbf{Y} \\
&- 2 \mathbf{F}^\top \mathbf{U}^\top \mathbf{U} \mathbf{Y})
\end{aligned}
\end{equation}

Therefore, the target label matrix $\mathbf{F}$ can be obtained as:
\begin{equation}
\mathbf{F} = \lambda(\mathbf{U}^\top - \mathbf{U}^\top \mathbf{\Theta}_{\mathbf{U}} \mathbf{U} + \lambda \mathbf{U}^\top \mathbf{U})^{-1} \mathbf{U}^\top \mathbf{U} \mathbf{Y}
\end{equation}

With the generated label matrix $\mathbf{F} \in \mathbb{R}^{n \times K}$ ($K = 2$ in our task), the new coming testing case can be identified as COVID-19 or CAP accordingly.

\section{Experiments}

\begin{table}[]
\caption{The definition of the confusion matrix for COVID-19 identification.}
\LARGE
\resizebox{\columnwidth}{!}{
\begin{tabular}{c|cc}
\hline
\textbf{} & \textbf{Classify as COVID-19} & \textbf{Classify as CAP} \\ \hline
\textbf{COVID-19} & True Positive (\textit{TP}) & False Negative (\textit{FN}) \\
\textbf{CAP} & False Positive (\textit{FP}) & True Negative (\textit{TN}) \\ \hline
\end{tabular}
}
\label{tab:confuse}
\end{table}

% --------------------main experiments-------------------
\begin{table*}
\centering
\caption{Performance comparison of different methods on the pneumonia dataset. 
(``${\dagger}$'' denotes the significance testing, $p-value < 0.05$.)}
\LARGE
\resizebox{\textwidth}{!}{
% \begin{tabular}{lc|cc|cc|cc|cc|cc|cc}
% \hline
% \multicolumn{2}{c|}{\textbf{Methods}} & \multicolumn{2}{c|}{\textbf{ACC}} & \multicolumn{2}{c|}{\textbf{SEN}} & \multicolumn{2}{c|}{\textbf{SPEC}} & \multicolumn{2}{c|}{\textbf{BAC}} & \multicolumn{2}{c|}{\textbf{PPV}} & \multicolumn{2}{c}{\textbf{NPV}} \\ \hline
% \textbf{SVM} & (\textit{p-value}) & 0.85884 & \textit{8.3824e-4} & 0.88902 & \textit{5.7014e-3} & 0.80999 & \textit{9.8357e-6} & 0.84950 & \textit{6.3223e-6} & 0.88219 & \textit{8.2747e-4} & 0.82148 & \textit{4.1927e-4} \\
% \textbf{MLP} & (\textit{p-value}) & 0.86888 & \textit{1.3914e-4} & 0.88503 & \textit{8.1923e-4} & 0.84245 & \textit{7.3121e-6} & 0.86374 & \textit{9.2958e-6} & 0.90191 & \textit{1.5186e-3} & 0.81741 & \textit{7.0136e-4} \\
% \textbf{iHL} & (\textit{p-value}) & 0.87299 & \textit{9.1836e-4} & 0.91682 & \textit{0.0538} & 0.80193 & \textit{8.8322e-7} & 0.85938 & \textit{2.6611e-6} & 0.88114 & \textit{1.9356e-7} & 0.85934 & \textit{0.0546} \\
% \textbf{tHL} & (\textit{p-value}) & 0.87448 & \textit{4.4017e-3} & 0.89397 & \textit{0.0031} & 0.84280 & \textit{1.9258e-5} & 0.86838 & \textit{2.0182e-4} & 0.90114 & \textit{2.0127e-3} & 0.83262 & \textit{3.3156e-3} \\
% \textbf{UVHL} & (\textit{std}) &
% \textbf{0.90018}$^{\dagger}$ & \textit{$\pm 0.0223$} &
% \textbf{0.91909}$^{\dagger}$ & \textit{$\pm 0.0291$} & 
% \textbf{0.87259}$^{\dagger}$ & \textit{$\pm 0.0274$} & 
% \textbf{0.89584}$^{\dagger}$ & \textit{$\pm 0.0210$} & 
% \textbf{0.91958}$^{\dagger}$ & \textit{$\pm 0.0222$} & 
% \textbf{0.86851}$^{\dagger}$ & \textit{$\pm 0.0483$} \\ \hline
% \end{tabular}}
\begin{tabular}{lc|cc|cc|cc|cc|cc|cc}
\hline
\multicolumn{2}{c|}{\textbf{Methods}} & \multicolumn{2}{c|}{\textbf{ACC}} & \multicolumn{2}{c|}{\textbf{SEN}} & \multicolumn{2}{c|}{\textbf{SPEC}} & \multicolumn{2}{c|}{\textbf{BAC}} & \multicolumn{2}{c|}{\textbf{PPV}} & \multicolumn{2}{c}{\textbf{NPV}} \\ \hline
\textbf{SVM} & (\textit{p-value}) & 0.84084 & \textit{1.173e-7} & 0.85714 & \textit{1.438e-6} & 0.81034 & \textit{4.235e-3} & 0.83374 & \textit{1.037e-4} & 0.89423 & \textit{0.0498} & 0.75200 & \textit{3.283e-6} \\
\textbf{MLP} & (\textit{p-value}) & 0.84685 & \textit{4.917e-6} & 0.86175 & \textit{1.082e-5} & 0.81897 & \textit{0.0153} & 0.84036 & \textit{2.349e-3} & 0.89904 & \textit{0.0507} & 0.76000 & \textit{8.777e-9} \\
\textbf{iHL} & (\textit{p-value}) & 0.85135 & \textit{5.260e-7} & 0.86327 & \textit{3.415e-4} & 0.83052 & \textit{0.0332} & 0.84790 & \textit{7.905e-3} & 0.90256 & \textit{0.2367} & 0.76866 & \textit{2.088e-8} \\
\textbf{tHL} & (\textit{p-value}) & 0.86486 & \textit{3.533e-4} & 0.89191 & \textit{2.851e-4} & 0.81743 & \textit{4.559e-3} & 0.85467 & \textit{0.0197} & 0.89898 & \textit{0.2383} & 0.80547 & \textit{7.071e-5} \\
\textbf{UVHL} & (\textit{std}) &
\textbf{0.89790}$^{\dagger}$ & \textit{$\pm 0.0223$} &
\textbf{0.93269}$^{\dagger}$ & \textit{$\pm 0.0291$} & 
\textbf{0.84000}$^{\dagger}$ & \textit{$\pm 0.0274$} & 
\textbf{0.88635}$^{\dagger}$ & \textit{$\pm 0.0210$} & 
\textbf{0.90654} & \textit{$\pm 0.0222$} & 
\textbf{0.88235}$^{\dagger}$ & \textit{$\pm 0.0383$} \\ \hline
\end{tabular}}

\label{tab:main_experiment}
\end{table*}
% --------------------main experiments-------------------

\subsection{Evaluation Metrics}
In our experiments, six criteria are employed to evaluate the COVID-19 identification performance, and the definition of the confusion matrix is shown in Table~\ref{tab:confuse}.

\begin{enumerate}

\item Accuracy (\textit{ACC}): \textit{ACC} measures the proportion of samples that are correctly classified.
$ACC = \frac{TP+TN}{TP+TN+FP+FN}$.

\item Sensitivity (\textit{SEN}): \textit{SEN} measures the proportion of actual positives that are correctly identified as such.
This metric is also called as ``recall'', reflecting the misdiagnose proportion.
In actual medical diagnostic application scenarios, this evaluation metric is more critical.
$SEN = \frac{TP}{TP+FN}$.

\item Specificity (\textit{SPEC}): \textit{SPEC} measures the proportion of actual negatives that are correctly identified as such.
It stands for the omission diagnose rate.
$SPEC = \frac{TN}{TN+FP}$.

\item Balance (\textit{BAC}): \textit{BAC} is the mean value of \textit{SEN} and \textit{SPEC}.
$BAC = \frac{SEN+SPEC}{2}$.

\item Positive Predictive Value (\textit{PPV}): \textit{PPV} measures the proportion of detected positives that are true positive.
$PPV = \frac{TP}{TP+FP}$.

\item Negative Predictive Value (\textit{NPV}): \textit{NPV} measures the proportion of detected negatives that are true negative.
$NPV = \frac{TN}{TN+FN}$.

\end{enumerate}

\subsection{Compared Methods}
The following popular classification approaches are used for comparison:
\begin{itemize}

\item Support Vector Machine (SVM) \citep{cortes1995support}:
It is a non-probabilistic linear classifier, used to perform supervised learning.
It selects a group of the training data as support vectors to determine the boundary that divides different categories apart as unambiguously as possible.
% It can be very stable in most classification tasks, especially in binary classification tasks. But as a linear classifier, SVM lacks the ability to mine correlation among features.

\item Multilayer Perceptron (MLP) Neural Network:
As the fundamental feed-forward artificial neural network, MLP can be utilized to perform binary classification with the \textit{cross-entropy} as the loss function.
% MLP can perform well in most tasks.
% However, as a deep neural network, MLP always requires a large quantity of data, which limits its performance on small dataset.

\item Inductive Hypergraph Learning (iHL) \citep{zhang2018inductive}:
In iHL, all available features are combined into one single feature, and then a projection is learned on the hypergraph structure, which is used to conduct classification task on the pneumonia instances.
This model learns the high-order representations from the training set and is evaluated in the testing set.

\item Transductive Hypergraph Learning (tHL) \citep{zhou2007learning}:
The transductive learning on hypergraph is conducted to learn the label matrix.
Both the training data and all testing data are employed in the hypergraph structure, yet leading to the commonly used semi-supervised learning approach.
\end{itemize}

% --------------------boxplot-------------------
\begin{figure}[ht]
    \begin{center}
    \includegraphics[width=\linewidth]{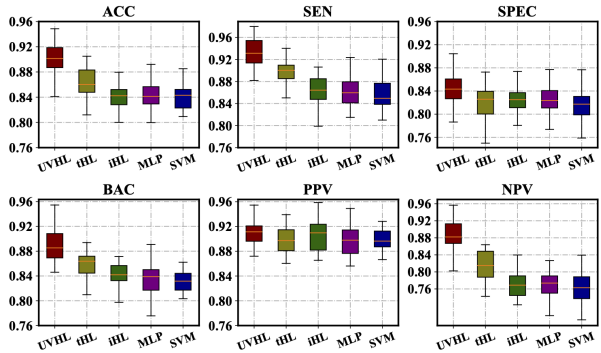}
    \end{center}
    \caption{\label{fig:boxline} The performance of UVHL and compared methods. The results show that UVHL outperforms other methods for all metrics.}
\end{figure}

\subsection{Implementation}
In our experiments, the whole dataset consists of 2,148 COVID-19 cases and 1,182 CAP cases.

We randomly divide them into 10 subsets and perform 10-fold cross-validation, in which 9 subfolds are used for training and the rest one is used for testing each time.
The data splitting process repeats 10 times, and the mean and standard deviation of all 10 runs are reported as the final result for comparison.
All features are normalized into $[0,1]$ in the training dataset, and the offset mean and variance are applied to the testing dataset for data normalization, respectively. 

For our UVHL model, K nearest neighbors are connected for each vertex when generating hyperedges.
We note that it is important to generate a suitable hypergraph structure for representation learning.
However, how to select the best $\mathcal{K}$ value in this procedure is difficult.
A large $K$ will lead to high dissimilarity insider the hyperedge, while a small $K$ may be not informative enough to the overall hypergraph structure.
To select a suitable $\mathcal{K}$, the following strategy is adopted to select $\mathcal{K}$.
First, a pool of candidate $\mathcal{K}$ values is set as $[2,3,...,20]$ in our experiments.
Given a set of training data and corresponding testing data, we further split the training data into 5 folds.
The 5-fold cross-validation is conducted on the training data, where different $\mathcal{K}$ are used.
We then collect the performance of different $\mathcal{K}$ on the training data, and the $\mathcal{K}$ with the best performance is used for testing.
In this way, the selection of $\mathcal{K}$ can be fully automatic and optimized.

\subsection{Results and Discussions}

Experimental results are demonstrated in Fig.~\ref{fig:boxline}, and the detailed mean value and the significance of the \textit{t}-test between UVHL and other methods are listed in Table~\ref{tab:main_experiment}.
From these results, we have the following observations:
\begin{enumerate}
    \item Our proposed method UVHL achieves the most reliable performance among all metrics. Compared with SVM and MLP, our approach obtains better performance (\ie, 6.79\% and 6.03\% relative improvement in terms of ACC, respectively), demonstrating that the hypergraph based approach has the effective ability to tackle the pneumonia identification task.
    \item Compared with other hypergraph based methods, \ie, inductive hypergraph learning (iHL) \citep{zhang2018inductive} and transductive hypergraph learning (tHL) \citep{zhou2007learning}, our approach achieves relative gains of 5.47\% and 3.82\% in terms of \textit{ACC}, respectively.
    \item Besides the better sensitivity value, our proposed UVHL method achieves much higher specificity value compared with all other methods. This indicates that our proposed method can \textit{not only} have high recall of COVID-19 patients \textit{but also} be effective on filtering CAP patients, which is quite useful in practice.
\end{enumerate}

\begin{table}[h]
\caption{Experimental comparison on the data uncertainty measurement.}
\LARGE
\resizebox{\columnwidth}{!}{
\begin{tabular}{c|c|cccccc}
\hline
& \textbf{Weighting strategy} & \textbf{ACC} & \textbf{SEN} & \textbf{SPEC} & \textbf{BAC} & \textbf{PPV} & \textbf{NPV} \\ \hline
\textbf{1} & \textbf{Equal Weight} & 0.85586 & 0.88426 & 0.80342 & 0.84384 & 0.89252 & 0.789912 \\
\textbf{2} & \textbf{Support Vectors} & 0.86066 & 0.87021 & 0.84442 & 0.85731 & 0.90983 & 0.78137 \\
\textbf{3} & \textbf{Aleatoric Uncertainty} & 0.87387 & 0.918919 & 0.78378 & 0.85135 & 0.89474 & 0.82857 \\
\textbf{4} & \textbf{Epistemic Uncertainty} & 0.88589 & 0.90741 & \textbf{0.84615} & 0.87678 & \textbf{0.91589} & 0.83193 \\
\textbf{5} & \textbf{Proposed Uncertainty} & \textbf{0.89790} &\textbf{0.93269} & 0.84000 & \textbf{0.88635} & 0.90654 & \textbf{0.88235} \\ \hline 
\end{tabular}
}
\label{tab:uncertainty_exp}
\end{table}

\subsection{Data Uncertainty Study}
To evaluate the effectiveness of our proposed data uncertainty method, we further conduct ablation experiments to compare variants of the data uncertainty measurement procedure.
First, we remove the uncertainty measurement procedure and treat all cases equally.
Secondly, the SVM-based uncertainty score is calculated, instead of that of using MLP.
Then, the two uncertainty measurements are used individually for comparison.
Experimental results are reported in Table~\ref{tab:uncertainty_exp}, from which we can have the following observations:
\begin{enumerate}
    \item Compared with the method without uncertainty, \ie, with equal weights, all the other methods with uncertainty can achieve better performance.
    \item The method with uncertainty from SVM performs worse than that of using MLP. It indicates that MLP has better identification effectiveness compared with SVM on uncertainty measurement. 
    \item Compared with the case of using aleatoric uncertainty and epistemic uncertainty individually, the use of both uncertainties, \ie, the proposed method, achieves the best performance, which demonstrates the effectiveness of our proposed data uncertainty strategy.
\end{enumerate}

\subsection{Analysis On Feature Types}
In this study, there are two types of features from CT, \ie, regional features and radiomics features.
Here, we evaluate the effectiveness of these two features on the task of COVID-19 identification. 
We have conducted experiments with our proposed method using each feature individually.
Experimental comparison is demonstrated in Table~\ref{tab:features_exp}.
Our method using regional feature has higher sensitivity, while the specificity is relatively lower, compared with the cases of using radiomics features.
These results indicate that regional feature is better in finding the true positive COVID-19 cases, while radiomics features have the advantage of identifying CAP cases.
When using both types of features in our proposed method, the performance becomes stable, along with both increasing sensitivity and specificity, as shown in the last row of Table~\ref{tab:features_exp}.
This observation demonstrates that our proposed method has the ability of jointly utilizing multi-type features and achieve better performance.

% ---------------------- Figure -----------------------
\begin{figure*}
    \begin{center}
    \includegraphics[width=\linewidth]{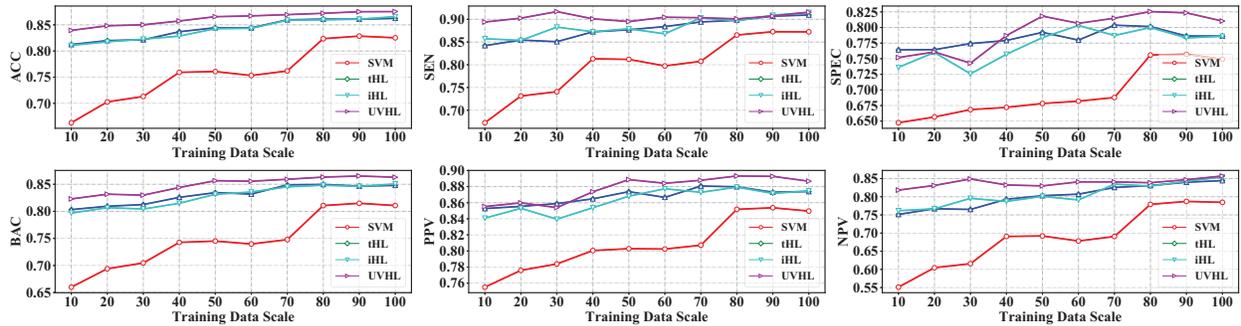}
    \end{center}
    \caption{\label{fig:line_graph} Performance comparison with respect to different scales of training data.}
\end{figure*}
% ---------------------- Figure -----------------------

\begin{table}[h]
\caption{Experimental comparison on different feature types and their combination.}
\LARGE
\resizebox{\columnwidth}{!}{
\begin{tabular}{c|cccccc}
\hline
\textbf{Feature Types} & \textbf{ACC} & \textbf{SEN} & \textbf{SPEC} & \textbf{BAC} & \textbf{PPV} & \textbf{NPV} \\ \hline
\textbf{Regional} & 0.85886 & 0.90323 & 0.77586 & 0.83954 & 0.88288 & 0.81081 \\
\textbf{Radiomics} & 0.85946 & 0.86982 & \textbf{0.84182} & 0.85582 & \textbf{0.90889} & 0.78012 \\
\textbf{Both} & \textbf{0.89790} &\textbf{0.93269} & 0.84000 & \textbf{0.88635} & 0.90654 & \textbf{0.88235} \\ \hline
\end{tabular}
}
\label{tab:features_exp}
\end{table}

\subsection{Analysis on Few Labeled Data}
\label{subsection: datascale}
As the large-scale labeled data for COVID-19 is expensive and maybe infeasible in emergent situations, how these methods perform with very limited labeled data is an important issue.
It should be noted that we have not included MLP, as MLP performs very badly when having very few training data.
To do that, we investigate how the compared methods work with respect to a small scale of labeled data from 10 to 100 for COVID-19 and CAP respectively.
In these experiments, 100 cases for each category are selected as the validation data.
The training data selection process repeats 10 times and the average performance is calculated for comparison.
Experimental results are shown in Fig.~\ref{fig:line_graph}.
As shown in these results, we can observe that SVM performs inferior in all settings when given just very few labeled data, and the hypergraph based methods perform the best. 
We can also observe that our proposed method, \ie, UVHL, can achieve very stable performance when only a few labeled data are available, which justifies the effectiveness of our proposed method in these difficult situations.

\section{Conclusion}
In this paper, we propose an uncertainty vertex-weighted hypergraph learning method to identify COVID-19 from CAP using CT images.
Confronting the challenging issues from noisy data and confusing cases with similar clinical manifestations and imaging features, our proposed method employs a hypergraph structure to formulate the data correlation among the known COVID-19 cases, the known as CAP cases, and the testing cases.
Through this method, two types of CT image features (including regional features and radiomics features) are extracted for patient representation.
To overcome the limitations of the noisy data, a data uncertainty measurement process is conducted to measure the uncertainty of each training case.
Finally, a vertex-weighted hypergraph learning process is used to predict whether a new case is COVID-19 or CAP.
We have conducted experiments on a large multi-center pneumonia dataset, including 2,148 COVID-19 cases and 1,182 CAP cases from 5 hospitals, and the experimental results demonstrate the effectiveness of our proposed method on identification of COVID-19 in comparison to the existing state-of-the-art methods.

In future work, the effectiveness of each individual feature should be fully investigated.
Regarding the limited data and possible evolution of COVID-19, it is important to explore small sample learning methods as well as transfer learning techniques on this difficult task of identifying COVID-19.

% \section*{Acknowledgments}
% This work was supported in part by the 
% National Natural Science Funds of China (61671267, 81871337),
% Beijing Natural Science Foundation (4182022),
% National Key Research and Development Program of China (2018YFC0116400),
% Wuhan Science and technology program (Grant No.2018060401011326), 
% Hubei Provincial Novel Pneumonia Emergency Science and Technology Project (2020FCA021),
% Huazhong University of Science and Technology Novel Coronavirus Pneumonia Emergency Science and Technology Project (2020kfyXGYJ014).

\bibliographystyle{model2-names.bst}\biboptions{authoryear}
\bibliography{refs}

% \section*{Supplementary Material}

% Supplementary material that may be helpful in the review process should
% be prepared and provided as a separate electronic file. That file can
% then be transformed into PDF format and submitted along with the
% manuscript and graphic files to the appropriate editorial office.

\end{document}